\documentclass{aa}  
\usepackage{graphicx}
\usepackage{txfonts}
\usepackage{xcolor}
\usepackage{alltt}
\begin{document}

   \title{Modeling and removal of optical ghosts in the PROBA-3/ASPIICS externally occulted solar coronagraph}

   \titlerunning{Optical ghosts in PROBA-3/ASPIICS}

   \author{S.~V.~Shestov\inst{1}\fnmsep\inst{2}  \and  A.~N.~Zhukov\inst{1}\fnmsep\inst{3} \and D.~B.~Seaton\inst{4}\fnmsep\inst{5}
          }

   \institute{Solar-Terrestrial Centre of Excellence --- SIDC, Royal Observatory of Belgium, 
              Avenue Circulaire 3, B-1180, Brussels, Belgium;  \email{s.shestov@oma.be}
	  \and
             Lebedev Physical Institute, Leninskii prospekt, 53, 119991, Moscow, Russia 
	  \and
	     Skobeltsyn Institute of Nuclear Physics, Moscow State University, Leninskie gory, 119991, Moscow, Russia 
          \and
	     Cooperative Institute for Research in Environmental Sciences, University of Colorado at Boulder, Boulder, CO 80305, USA 
	  \and
	     National Centers for Environmental Information, National Oceanic and Atmospheric Administration, Boulder, CO 80305, USA; }

   \date{Received November 6th, 2018; accepted December 8th, 2018}
   \abstract
      {ASPIICS is a novel externally occulted solar coronagraph, which will be launched onboard the PROBA-3 mission of the European Space Agency. The
	external occulter will be placed on the first satellite $\sim 150$~m ahead of the second satellite that will carry an optical
	instrument. During 6 hours per orbit, the satellites will fly in a precise formation, constituting a giant externally occulted
	coronagraph. Large distance between the external occulter and the primary objective will allow observations of the white-light solar
	corona starting from extremely low heights $\sim 1.1\mathrm{R}_\sun$.}
      {To analyze influence of optical ghost images formed inside the telescope and develop an algorithm for their removal.} 
      {We implement the optical layout of ASPIICS in Zemax and study the ghost behaviour in sequential and non-sequential regimes. We identify
        sources of the ghost contributions and analyze their geometrical behaviour. Finally we develop a mathematical model and software to
        calculate ghost images for any given input image.}
      {We show that ghost light can be important in the outer part of the field of view, where the coronal signal is weak, since the energy of
	bright inner corona is redistributed to the outer corona. However the model allows to remove the ghost contribution. Due to a large
	distance between the external occulter and the primary objective, the primary objective does not produce a significant ghost. The
	use of the Lyot spot in ASPIICS is not necessary.}
      {}
   \maketitle

\section{Introduction}
  The difficulty of observing the low solar corona in white-light coronagraphic observations is caused by the two factors: the very high dynamic
  range of the corona and the presence of intense light from the solar disk diffracted by the external occulter (EO)
  \citep{Koutchmy1988,Shestov2018}.

  The high dynamic range of coronal brightness, which may vary from $10^{-5}$ of mean solar brightness (MSB) at $1.01\mathrm{R}_\sun$ to less than
  $10^{-10}$~MSB at $3\mathrm{R}_\sun$, makes instrumentation intended to observe this region in visible light very sensitive to the presence of stray
  light of any nature, like diffraction, scattering, ghost light. Even a small fraction of light from the inner corona being scattered to higher 
  altitudes can significantly modify the observed brightness and make the interpretation of results difficult. The contribution of stray light
  complicates diagnostics of plasma temperature and electron density in both coronagraphic \citep{Wang2017}, imaging
  \citep{2012ApJ...749L...8S,0004-637X-781-2-100}, spectroscopic \citep{2008A&A...483..271D,2011ApJ...736..101H,2018ApJ...856...28W}
  and even radio \citep{1974A&AS...15..417H,1985A&A...153..139W} observations. Besides, the presence of stray light
  in coronal images also reduces contrast and thus precludes small-scale features from being identified. 

  For white-light coronagraphic observations diffraction plays an especially important role in stray light generation. In internally
  occulted coronagraphs the light diffracted by the entrance aperture produces an intense, sharply decreasing stray light pattern on the
  detector.  The intensity of the diffraction in the region of the inner corona becomes greater than the intensity of the corona itself
  \citep{2017A&A...599A...2R}.  Furthermore, the situation is worsened by the scattering of direct sunlight on the primary objective
  \citep[e.g.][]{Thompson2010}.  Externally occulted coronagraphs place the primary objective (PO) in the umbra of the external occulter, which
  alleviates the problem of scattering, but this provides no improvement of diffraction in the inner-most corona \citep[see Fig.~10
  in][]{2017A&A...599A...2R}. The remaining diffraction is removed by the internal occulter (IO), which is one of the key
  optical elements of externally occulted coronagraphs. The IO is the conjugated element to the EO with respect to the PO, it is slightly
  oversized in order to block the diffraction ring produced by the EO. Finally, on the detector the diffracted light has the shape
  of a bright ring with the angular size of the IO, and its full intensity is determined primarily by the IO oversizing.
  The intensity of diffraction in externally occulted coronagraphs is usually smaller than in internally occulted ones, however external
  occulters produce significant vignetting of the inner field of view (FOV) \citep{Koutchmy1988,2011ExA....29..145B,2017A&A...599A...2R}.
  This explains the relatively high innermost observational heights of white-light coronagraphs. 

  ASPIICS (Association of Spacecraft for Polarimetric and Imaging Investigation of the Corona of the Sun) is a novel white-light solar
  coronagraph that will make regular observations of corona within the FOV from $\sim 1.08\mathrm{R}_\sun$ up to $3.0 \mathrm{R}_\sun$
  \citep{2010SPIE.7731E..18L, 2015SPIE.9604E..0AR,doi:10.1117/12.2312493}. This will be possible owing to the European Space Agency's (ESA)
  formation flying (FF) mission PROBA-3 (Project for On-Board Autonomy, \citealt{Bernaerts2002}), in which the external occulter will be
  placed on the first satellite, and the telescope will be placed on the second \citep{2005SPIE.5899..221L,2006SPIE.6265E..24V}. 

  Good performance of white-light coronagraphs requires the reduction of stray light. This usually includes the
  optimization of the shape of the external occulter, as in LASCO C2 \citep{Bout2000} or METIS on board Solar Orbiter
  \citep{doi:10.1117/12.2309079}, the introduction of specialized diaphragms like the Lyot spot in
  LASCO C2 \citep{Brueckner1995, Llebaria2004}, the utilization of super-polished lenses and mirrors like in METIS \citep{Sandri2018},
  and COR1 \citep{Thompson2003}, and the use of a set of baffles like in HI-1 and HI-2 onboard STEREO \citep{Eyles2009} etc.

  In ASPIICS, various means are also used in order to reduce the level of stray light. The EO edge will have a toroidal shape, which will
  reduce the intensity of the diffracted light entering the telescope by a factor of two \citep{Landini2010,doi:10.1117/12.2313258};
  technological limitations do not allow the use of superior EOs like multi-disk, serrated, threaded cone or even petal-shaped (see
  \citealt{Bout2000} for a review of occulter systems, and the recent analytical/numerical analysis of serrated occulters by
  \citealt{Rougeot2018}). The lenses will have high quality surfaces with microroughness better than $\sim 1$~nm and will be covered with
  anti-reflection (AR) coatings with $R\sim0.3$\%; the size of the IO will be chosen to achieve balance between the level of diffracted
  light and minimal observational height \citep{Shestov2018}. On-ground data processing can yield further improvement of registered data.
  For example, special analyses have been undertaken to model stray light in COR1 \citep{Thompson2010}, LASCO~C2 \citep{Llebaria2012}, HI-1
  \citep{Halain2011}.
  
  The expected lifetime of ASPIICS requires thorough analysis of various stray light contributions and the algorithms for their
  removal.  Preliminary comparison of various sources of stray light in ASPIICS showed that the most significant contribution is due to ghost
  images, which are formed by the backreflection of the bright inner corona from the detector and neighboring optical surfaces. 

  The aim of the present paper is to analyse the behavior of optical ghost images, and compare their contribution with other stray light sources.
  We develop a geometrical/mathematical model of ghost images and provide an algorithm/software for the removal of the effect. 

  The paper is structured as follows: in Sect.~\ref{sec2} we describe the optical layout of ASPIICS, in Sect.~\ref{sources-sec} we
  discuss possible sources of stray light, in Sect.~\ref{ghosts-sec} we consider the mechanism of formation of ghost images and
  their characteristics. In Sect.~\ref{realistic-sec} we analyze ghost images for realistic coronal scenes (both synthetic and observed during
  a total eclipse) and discuss a procedure for the ghost light removal. In Sect.~\ref{comparison-sec} we compare the contributions of ghost images
  and other sources of stray light. In Sect.~\ref{discussion-sec} we discuss our results and draw conclusions.

\section{Optical layout}
\label{sec2}
  Below we give a description of two models: a simplified one that shows functional purposes of optical elements, and a full
  one, which provides further details.

\subsection{Functional optical layout}
\begin{figure*}
    \centering
    \includegraphics[width=18cm]{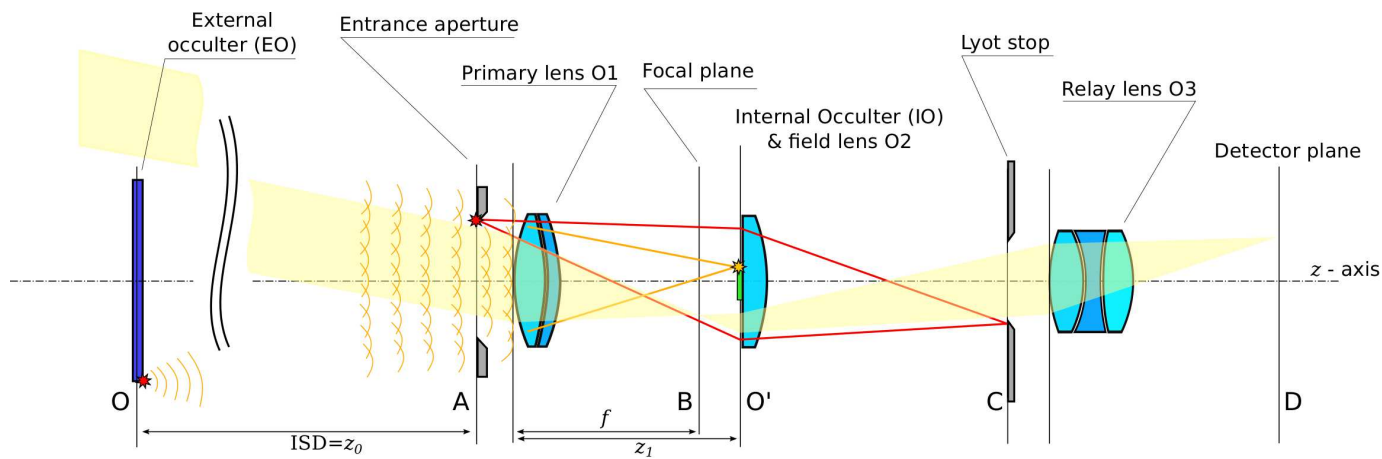}
    \caption{Functional optical layout of the ASPIICS coronagraph. Dim yellow light represents regular coronal beam, orange
	    arches and lines -- light that diffracts on EO, the red lines represent the light diffracted at the entrance aperture.}
    \label{layout}
  \end{figure*}

  The functional optical layout of the ASPIICS coronagraph is given in Fig.~\ref{layout}.  The external occulter (plane $O$) with
  $\diameter1420$~mm is situated on the occulter satellite $z_0=144\,348$~mm ahead of the coronagraph satellite, which carries the optical
  instrument. The telescope consists of an entrance aperture with $\diameter50$~mm (plane $A$) and a primary lens O1 (a cemented doublet)
  with focal length $f\approx330.3$~mm that makes an image of the corona in the primary focal plane $B$.  The light diffracted at EO is
  focused in the $O'$ plane, which is the conjugate to the plane O. The $O'$ plane is situated at a distance $z_1$, or $\sim 0.76$~mm
  further than $B$. The major part of the diffracted light is cut out by the internal occulter, the radius of which must be carefully
  chosen to obtain a compromise between good rejection of diffraction, minimum observational height and vulnerability of the coronagraph to
  possible tilts and other misalignments \citep{2017A&A...599A...2R,Shestov2018}. The IO is deposited directly on the surface of the field
  lens O2 (a singlet lens).  The lens O2 along with O1 make an image of the entrance aperture on the $C$ plane, where the Lyot stop is
  placed. The Lyot stop is slightly undersized ($\sim 97\%$) with respect to the entrance aperture in order to reject the light diffracted
  at the entrance aperture.  Simultaneously, the O2 lens projects the entrance aperture to the relay lens O3, ensuring that the light
  propagates further into the coronagraph and justifying its name -- field lens.  Finally, both O2 and O3 project the primary focus $B$ onto
  the detector plane $D$, making the effective focal length $f_\mathrm{eff}\approx734.6$~mm. An extensive discussion of the functional
  optical layout is given in \citet{2017A&A...599A...2R,Shestov2018}.  

  In Fig.~\ref{layout} dim yellow light represents regular coronal beam, orange arches represent light that diffracts on EO. After
  propagation through the entrance aperture and O1, this light is focused (orange lines) in the $O'$ plane. The red lines represent the
  light diffracted at the entrance aperture.

\subsection{Detailed optical layout}
\begin{figure*}
    \centering
    \includegraphics[width=17cm]{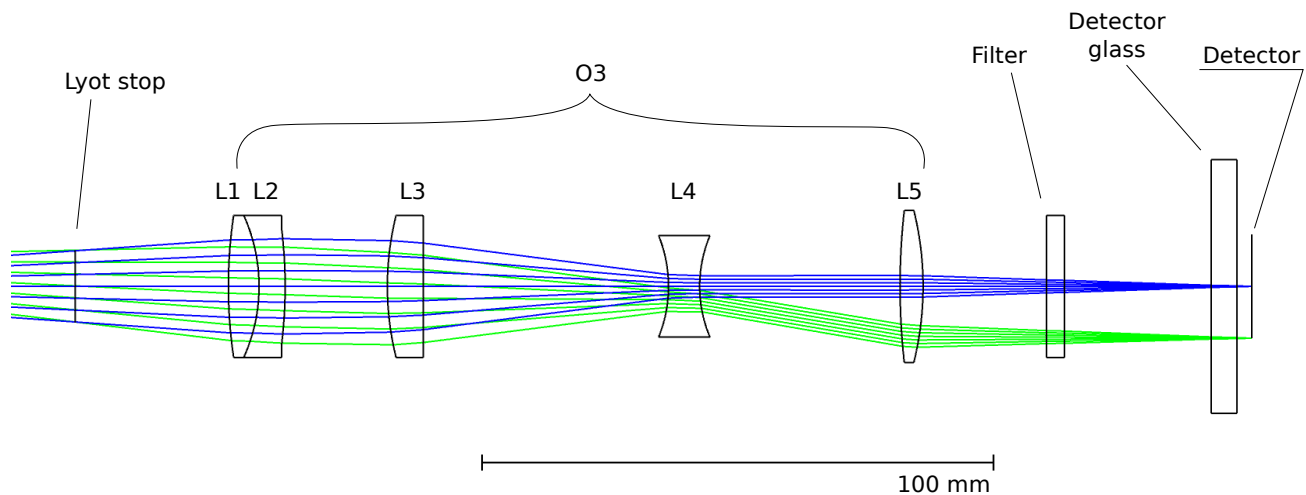}
    \caption{Detailed optical layout of the rear part of the ASPIICS coronagraph. Blue and green lines denote ray paths propagating along the
    optical axis (blue) and at the edge of the FOV -- $0.8^\circ$ ($3\mathrm{R}_\sun$) (green).}
    \label{layout-zoomed}
  \end{figure*}

  The real optical layout differs mainly in the more complicated relay lens O3, the presence of spectral filters (a filter wheel
  places one of the six filters into the beam) and the presence of the detector glass. Detailed description of the optical layout of ASPIICS is given in
  \citet{2015SPIE.9604E..0BG}. In Fig.~\ref{layout-zoomed} we provide the rear part of the optical layout of ASPIICS,  produced with Zemax
  OpticStudio.

  The detector is a front-side illuminated CMOS detector with $2048 \times 2048$ pixels, each pixel
  being $10 \times10$~$\mu$m. Its geometrical size determines the outer FOV of the coronagraph up to $3R_\sun=0.8^\circ$.

  The detector glass is a plane-parallel BK7 plate with a neutral-density 50\% filter. It is situated $\sim 5$~mm ahead of the detector chip
  and is used to reduce the relative contribution of the ghost light.
  
  The six filters are installed in the filter wheel, which can put any one of them into the optical beam. The filters are: one wideband filter
  for the spectral range 535--565~nm, two narrowband filters for 530.4~nm (Fe~XIV) and 587.7~nm (He~I), and three polarizers rotated by
  $60^\circ$ with respect to each other combined with the wideband spectral filters. Each filter is either a single plane-parallel glass
  or several glasses stacked together (like in the case with the polarizers) with AR coatings on the sides. In the current
  analysis we represent the filter as a single plane-parallel glass situated $\sim 40$~mm ahead of the detector. 

  The relay lens O3 consists of 5 individual lenses. The last lens -- O3/L5 is a telecentric lens, which re-images the entrance aperture
  to $z\approx-1\,260$~mm. This is done to improve performance of the spectral filters by reducing the range of incident angles both for an individual
  beam, and for the whole FOV.

\section{Sources of stray light}
  \label{sources-sec}
  Preliminary analysis of various stray light contributors \citep{Galy2018} demonstrated the significance of ghost light. The
  analysis was based on the raytracing approach and took into account such effects as ghost light and scattering on the lens
  surfaces. A thorough analysis of the diffracted light was performed by \citet{2017A&A...599A...2R} and \citet{Shestov2018}, thus we
  summarize their findings below.
  
  The relative importance of the ghost images is explained by the fact that the light from the inner-most -- and thus brightest -- corona is
  re-imaged by ghost reflections to higher heliocentric heights, i.e. towards the outer FOV. The intensity of the ghosts in the
  outer FOV regions can be as high as 10\% of the intensity of the local corona. To reduce the impact of this ghost light, an
  additional ND50\% was introduced in the detector glass. With the presence of the ND filter the regular coronal beam loses 50\% of energy,
  while the ghost light is decreased by factor 8 \citep{Galy2018}, passing through the filter two times more. 

  The light scattered on the lens surfaces provided a smaller contribution to the final stray light. The scattering has a higher efficiency at
  smaller angles and smaller efficiency at higher angles. Thus, scattering of the bright light from the inner corona to the greater heights
  was less efficient. Currently an advanced investigation of the lens scattering characteristics is being performed \citep{Rougeot2018b}. 

  Diffracted light has the shape of a very bright and defocused ring, the size and the position of which correspond to the edge of the IO
  projected to the detector plane \citep{2017A&A...599A...2R,Shestov2018}. Since the effect of diffraction consists in interference of
  wavefront regions propagating through the opening in the apertures (Huygens-Fresnel principle), the resulting diffraction weakly depends on the lens
  characteristics, such as geometry and roughness, which is opposite to the behaviour of ghost and scattered light. In the present analysis 
  diffracted light is considered as an additional light source, which enters the telescope along with the coronal light, and undergoes the
  same ghost reflections and scattering as the coronal light does. The amount of the diffracted light incident on the primary objective is
  significantly -- up to 3 orders of magnitude -- larger than the amount of diffracted light coming to the detector. This is explained by
  the fact that a major part of diffraction is blocked by the IO. Thus particular attention should be paid to the analysis of ghost
  reflections and scattering in the primary objective.

  Preliminary analysis demonstrated that the contribution of other stray light sources is negligible \citep{Galy2018}. 

\section{Ghost image formation}
  \label{ghosts-sec}
 
\subsection{Identification of the ghosts}
  \label{ghosts-basic-sec}
  In order to analyse properties of ghost images, we implemented an optical model of ASPIICS in a raytracing software package -- Zemax
  OpticStudio in a non-sequential regime (see description of various regimes of Zemax in Appendix~\ref{zemax}).  Since at this first stage
  of analysis we are interested in identification (determining of the parent surfaces) and general geometrical properties (size, location)
  of ghost images, we do not take into account any AR coatings. The reflection coefficients are calculated automatically based on glass
  properties and amounted to 4\% for lenses, while the detector reflectivity was taken to be 15\%. We also do not implement
  the IO in the optical layout; its effect may result in additional partial/full blocking of ghost images formed by the primary objective.

  We used two beams tilted at $0.28^\circ$ ($1.05 \mathrm{R}_\sun$) and $0.56^\circ$ ($2.1 \mathrm{R}_\sun$) as an
  input and calculated the image on the detector with ghost reflection enabled in Zemax. We analysed the image on the
  detector paying attention to the second order ghosts, i.e. the ghosts formed as a result of two reflections, which is justified by the
  small amount of the reflected light. Since in the non-sequential regime Zemax preserves the information about how a
  particular ray was formed, or which surfaces it passed through, we were able to reveal the ghost rays that were formed by individual
  surfaces. Thus we were able to identify the ghosts' origins, their geometry and spatial behaviour. 

\begin{figure*}
    \centering
    \includegraphics[width=14cm]{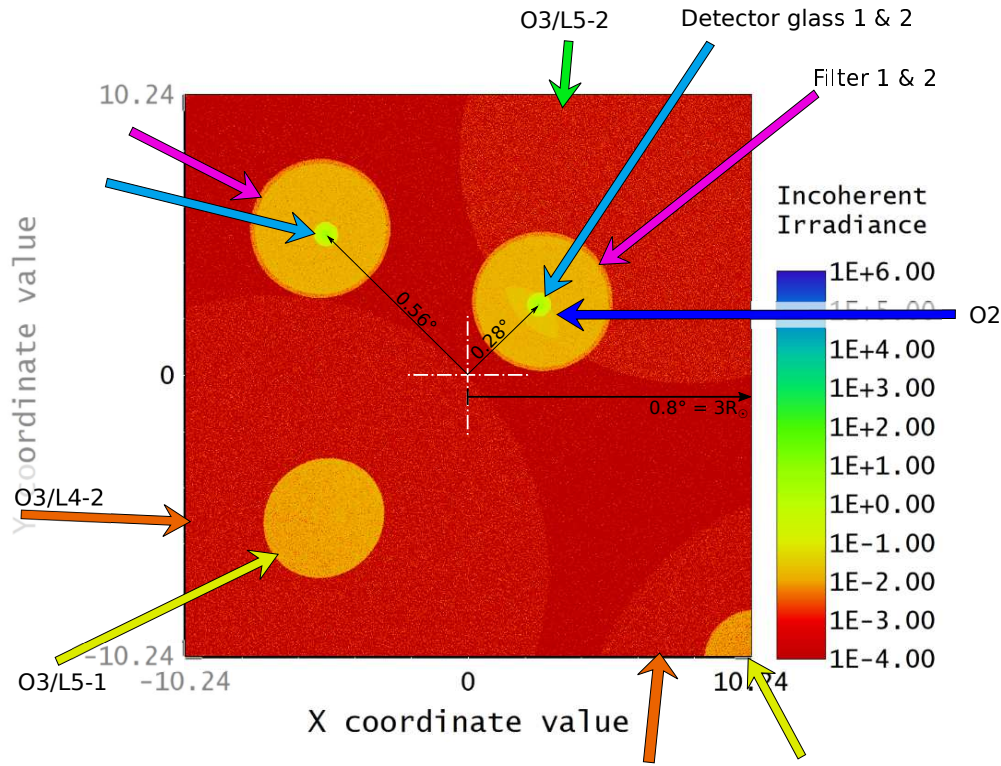}
    \caption{Image on the detector produced in Zemax OpticStudio in non-sequential regime with two input beams tilted at $0.28^\circ$ and
      $0.56^\circ$. The white cross denotes the optical axis (center of the FOV), the input beams are marked with thin black arrows, color
      arrows show ghost contributors produced by the backreflection from the detector and particular optical surfaces. Annotated color
      arrows correspond to the ghosts, produced by the $0.28^\circ$ input beam, and the rest correspond to the ghosts produced by the
      $0.56^\circ$ beam.}
    \label{ghost-nature}
  \end{figure*}

  A complex image showing two input beams and ghost images created due to backreflection from the detector and another optical surface is
  shown in Fig.~\ref{ghost-nature}. The positions of the input beams are marked with thin black arrows. The reflections from both
  surfaces of the detector glass generate tiny circular images (blue arrows) co-centered with the input beams. The reflections from the filter
  surfaces yield the larger circular images (magenta arrows) almost co-centered with initial beams.  The co-centering of these ghosts is due to
  the effect of the telecentric lens L5, which re-images the aperture almost to infinity, making the chief ray nearly perpendicular to the
  detector surface. The rear surface of the O3/L5 lens (O3/L5-2) deflects the beam and changes its convergence, thus its ghost image (green arrow) has
  a significantly larger diameter and is not co-centered with the input $0.28^\circ$ beam. For the $0.56^\circ$ beam this ghost falls
  outside the detector.  The front surface of the O3/L5 lens (O3/L5-1) also changes the convergence and deflects the beams symmetrically with respect
  to the optical axis.  Its ghosts are shown by yellow arrows in Fig.~\ref{ghost-nature}.  The rear surface of the O3/L4 lens produces
  deflected and enlarged images (orange arrows in Fig.~\ref{ghost-nature}), whereas the front surface of this lens produces a very large image
  uniformly covering the whole detector.  The O2 lens produces a complex image that resembles the images from the detector glass and a fish-like
  structure (a blue arrow), seen near the $0.28^\circ$ input beam. Most of these ghosts (beside the fish-like contribution of
  the O2) are not vignetted inside the telescope.

\begin{figure}
    \resizebox{\hsize}{!}{\includegraphics{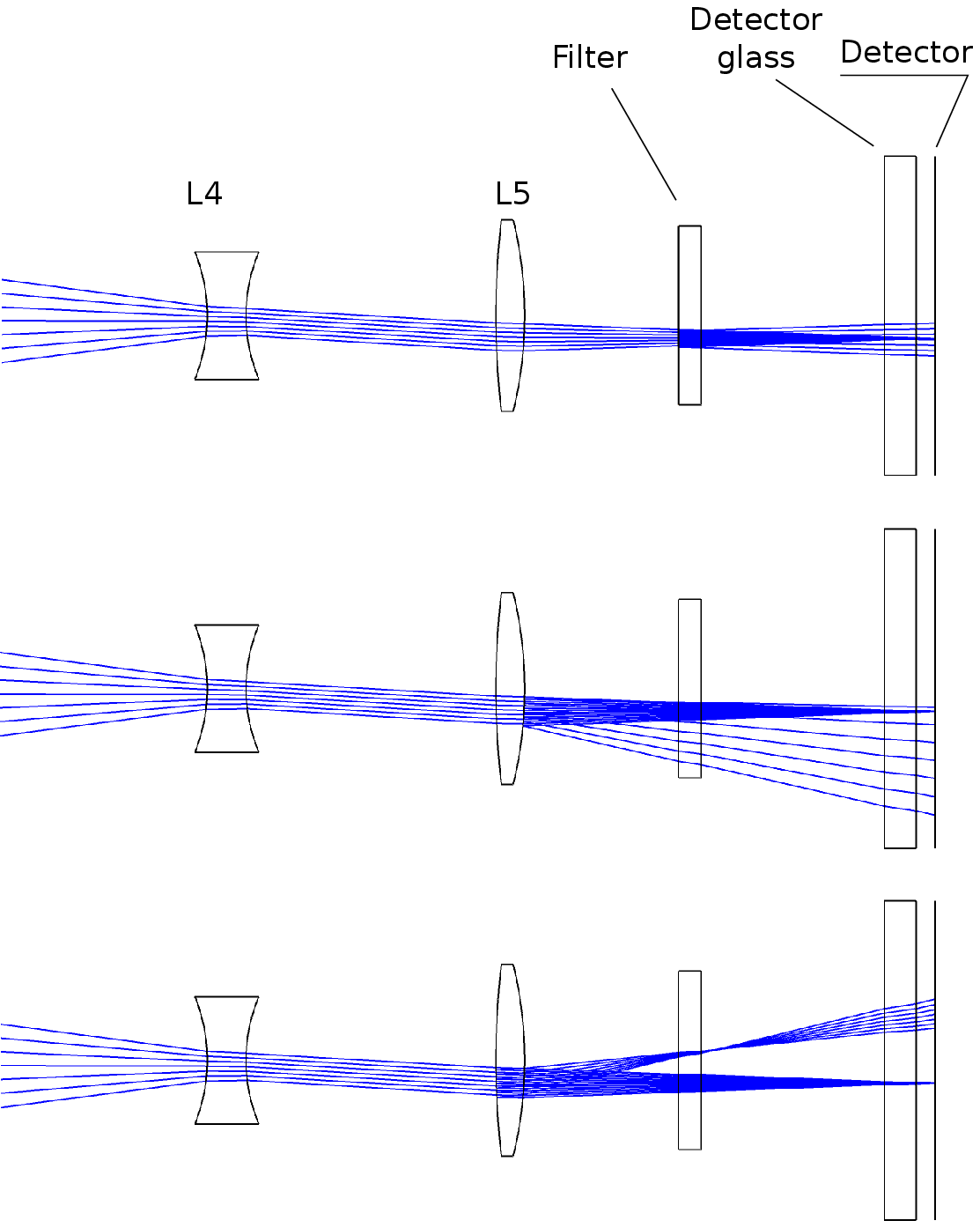}}
    \caption{Ray paths produced by an incoming beam and a ghost beam, formed by the detector and one of the optical surfaces. The incoming
      beam is backreflected from the detector, travels back almost the same way, experience ghost reflection and travels back forming
      different path. Top panel shows ray paths produced by the filter, middle panel -- by the rear surface of the O3/L5 lens, bottom panel
      -- by the front surface of the O3/L5 lens.} \label{ghost-sketch}
  \end{figure}

  In Fig.~\ref{ghost-sketch} we show ray paths produced by the front surface of the filter, the rear and front surfaces of the O3/L5 lens.
  The two bottom panels show the deflection and change of convergence by the two surfaces of the O3/L5 lens.

  We paid special attention to the ghosts produced by the primary objective. The O1/L1 and O1/L2 lenses produce extremely weak and uniform
  illumination on the detector. Obviously, this is explained by the ghost focus position far from the detector and the beam divergence, due
  to which only a small portion of rays ultimately reaches the detector. The total ghost energy produced by the O1 lens on the detector
  amounts to 0.03\% of the main ghost contributions. Such ghost behavior is in stark contrast with the LASCO C2 and LASCO C3 coronagraphs
  onboard SOHO \citep{Brueckner1995, Llebaria2004}, where an additional opaque disk called Lyot spot was placed in front of the
  relay lens. The Lyot spot prevented the ghost created in the primary objective from further propagation.  These ghosts were created in the
  primary objective by the light diffracted at the external occulter placed $\sim 820$~mm (in LASCO C2) ahead of the primary objective. In
  the case of ASPIICS, the EO is placed $\sim 150$~m ahead of the primary objective and the corresponding ghosts are formed between the
  primary objective and the field lens O2. The divergence of the ghost beams permits only a negligible part of the energy to propagate farther
  into the optical system. The possible influence of intense diffracted light falling on the primary objective with the energy $\times10^3$ higher
  than the energy of the diffracted light on the detector is alleviated by the extreme loss (factor $3\cdot10^{-4}$) of energy of
  the corresponding ghost.

  The total energy of the O2 contribution amounts only to 14\% of the detector backreflected ghosts, and since the relative
  intensity of the fish-like component (Fig.~\ref{ghost-nature}) is small, the contribution of the O2 lens can be included in the detector
  glass ghost energetics. The full ghost energy (with any second surface, not necessarily the detector) produced by the O1/L1, O1/L2, O3/L1,
  O3/L2 and O3/L3 lenses amounts to 0.01\%, 0.02\%, 0.14\%, 0.51\% and 0.9\%, respectively, of the total energy produced by the detector backreflections,
  thus these lenses are omitted from further consideration.  The relative contribution of other ghosts, such as higher order ghosts, or
  second order ghosts produced by O3/L4, O3/L5, the filter and the detector glass (i.e. excluding the detector backreflections) is also
  relatively small. In the current model, with no AR coatings, the intensity of the ghosts created by the detector backreflection amounts to
  61\% of all ghosts, and most of the remaining 39\% is due to ghost reflection by the detector glass and the filter. The introduction of a
  model AR coating with $R=0.5$\% reflectivity on several surfaces closest to the detector immediately increases this ratio to 90\% or better, 
  because the relative contribution of other ghosts decreases.  Thus we conclude that the presence of a high-reflectivity AR with
  $R\sim0.3$\% coating will make the contribution of the detector ghosts dominant. 
  
  Another effect to consider is that after reaching the detector all ghost light will be reflected once again, and after that will produce
  additional ghosts following similar geometrical rules. However, the intensity of these ghosts will be small because of the small value of the
  reflection coefficient. In Sect.~\ref{removal-sec} we will confirm the validity of this assumption.

  We conclude that the L4 and L5 lenses of O3, the filter and the detector glass are the primary contributors to ghost light.
  Their geometrical properties and energetics will be analyzed and modeled further. 

\subsection{Geometrical behaviour and energetics of the ghosts}
\begin{figure}
    \resizebox{\hsize}{!}{\includegraphics{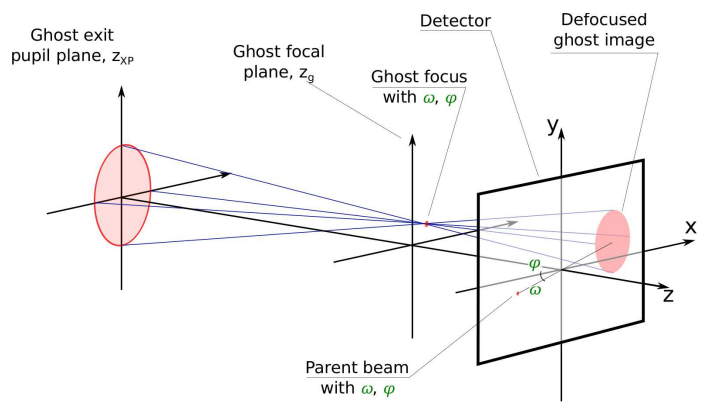}}
    \caption{A sketch representing geometry of the defocused image of a ghost. Configuration of the parent beam and the defocused image
    correspond to the O3/L5-1 lens surface.}
    \label{ghost-formation-geometry}
  \end{figure}

  In Sect.~\ref{ghosts-basic-sec} we used the non-sequential regime of OpticStudio to produce the ghost image for a given input image. In
  order to analyze the geometrical behavior of the selected ghosts depending on the incoming beam angle, we used the sequential
  regime of Zemax and retrieved first-order parameters of ghosts (which are obtained in a paraxial approximation, see
  Appendix~\ref{zemax}). Thus for each pair of optical surfaces we obtained the effective ghost focal distance $f_g$ and its
  working $f/\#$ ratio, the position of the ghost exit pupil $z_{XP}$, and the position of the ghost focal plane $z_g$. A sketch
  representing the formation of the defocused ghost image on the detector is shown in Fig.~\ref{ghost-formation-geometry}. The center of the
  ghost image is determined by the position of the ghost focus and the ghost exit pupil, while the diameter of the defocused image is
  determined by the working $f/\#$ ratio and distances between different planes. The main characteristics of the major ghosts are summarized
  in Table~\ref{geometry-table}.

  \begin{table*}
    \caption{Characteristics of the main ghosts, formed by the backreflection from detector and another optical surface.
    $z_{XP}$ -- exit pupil position, $z_g$ -- effective focus position, $f_g$ -- effective focal length, $f/\#$ -- working $f/\#$,
    $R_\mathrm{det}$ -- reflectance of the detector, $R$ -- reflectance of the second surface.  $z=0$ position corresponds to
    the detector plane, $z$-negative direction is towards the entrance aperture. Positive focal length corresponds to the situation where a
    ghost image is formed on the same side along with parent beam  with respect to the center of the optical axis. }
    \label{geometry-table}
    \begin{tabular}{l l l l l l l l}
      \hline \hline 
      Surface			     & Symbol  & $z_{XP}$, mm &
							   $z_g$, mm &
								     $f_g$, mm &
			 							 $f/\#$ & $R_\mathrm{det}$, \% & $R$, \% \\
      \hline
      Detector glass, front	     & G1      & -1273.72  & -12.19 & 731.88 & 14.64 & 15.0 & 0.3  \\
      Detector glass, rear	     & G2      & -1267.14  & -5.60  & 731.88 & 14.64 & 15.0 & 0.3  \\
      Filter glass, front	     & F1      & -1335.83  & -74.30 & 731.88 & 14.64 & 15.0 & 0.23 \\
      Filter glass, rear	     & F2      & -1331.22  & -69.69 & 731.88 & 14.64 & 15.0 & 0.23 \\
      Relay objective, L5 lens front & O3/L5-1 & -41.71	   & -30.46 & -385.98\tablefootmark{b} & 7.72 & 15.0  & 0.23 \\
      Relay objective, L5 lens rear  & O3/L5-2 & -92.22	   & -82.23 & 249.06 & 4.98  & 15.0 & 0.23 \\
      Relay objective, L4 lens rear  & O3/L4-2 & -132.01   & -114.02& -282.36\tablefootmark{b} & 5.65 & 15.0  & 0.23 \\
      \hline
    \end{tabular}
    \tablefoot{
      \tablefoottext{b}{negative value means symmetrically with respect to the optical axis;}
    }
  \end{table*}  

  The total energy stored in each ghost image is determined by the intensity of the parent beam $I$, the reflectivity of the detector ($\sim
  15\%$), the reflectivity of the particular optical surface (determined by the deposited AR coating with $R\sim0.3\%$), and the
  transmission of the glass elements (amounts to $\sim70$\% for all the lenses). The total energy of the ghost image, however, is uniformly spread
  across its full area. The specific intensity (per pixel) \texttt{di} of a particular ghost image depends on the parent surface: the specific intensity
  of the detector glass ghost will be considerably higher than the intensity of the O3/L5 lens ghost due to its considerably smaller area. We
  have verified that all ghosts considered here (beside the fish-like contributor of the O2) do not experience additional vignetting inside
  the telescope.

\subsection{Geometrical model and software for calculation}
  \label{model-sec}
  The computation of a ghost image using a raytracing approach is computationally rather expensive: even for a simple input image
  with just two incoming beams, each containing $\sim 10^5$ rays, it takes up to 10~min to compute the final image using OpticStudio on a contemporary
  Intel i5 computer. Additionally, there are various difficulties in embedding raytracing software into an automated data processing workflow.
  Thus we have developed software for calculating the ghost image for a given input image in order to embed it into the on-ground image
  processing pipeline.  
  
  We assume that every pixel of the input image represents an incoming plane-parallel wave and calculate the ghost response for every
  plane-parallel component. We loop through all the pixels of the input image using two nested loops in $y$ and $x$, and for every pixel we
  calculate its radial ($\omega$) and polar ($\varphi$) coordinates. Based on these coordinates we calculate the position of the ghost focus
  $x_g$ and $y_g$ in the ghost focal plane with coordinate $z_g$ (using the geometrical parameters from Table~\ref{geometry-table}). The
  ghost focus and the ghost exit pupil \texttt{XP} determine the position and the size of the defocused ghost image on the
  detector (see Fig.~\ref{ghost-formation-geometry}).  In two next-level loops we check whether every pixel on the detector 
  with the coordinates $\mathrm{[K,P]}$ falls within the defocused ghost image. We perform this by verifying if the line through the points
  $(x_g,y_g,z_g)$ and $(K,P,0)$ crosses the exit pupil. In case the pixel belongs to the defocused ghost image we increase the intensity in
  the ghost image $\mathrm{G[K,P]}$ by the specific intensity \texttt{di} (depends on the intensity of the initial pixel, reflectivities and the area
  of the ghost). The procedure is carried out for every pair of ghost-producing surfaces, i.e. for every row in the
  Table~\ref{geometry-table}, with the following algorithm: 
  
  \begin{alltt}
for y=1,2048
  for x=1,2048
    \(\omega = \arctan(\sqrt{x\sp{2}+y\sp{2}}/f) \)
    \(\varphi = \arctan(y/x) \)
    \(x\sb{g} = f\sb{g} \tan\omega \cos\varphi \)
    \(y\sb{g} = f\sb{g} \tan\omega \sin\varphi \)
    for P=1,2048
      for K=1,2048
        if line(\((x\sb{g},y\sb{g},z\sb{g})\leftrightarrow(K,P,0)\)) crosses XP 
        then G[K,P] = G[K,P]+di
  \end{alltt}

\begin{figure}
    \resizebox{\hsize}{!}{\includegraphics{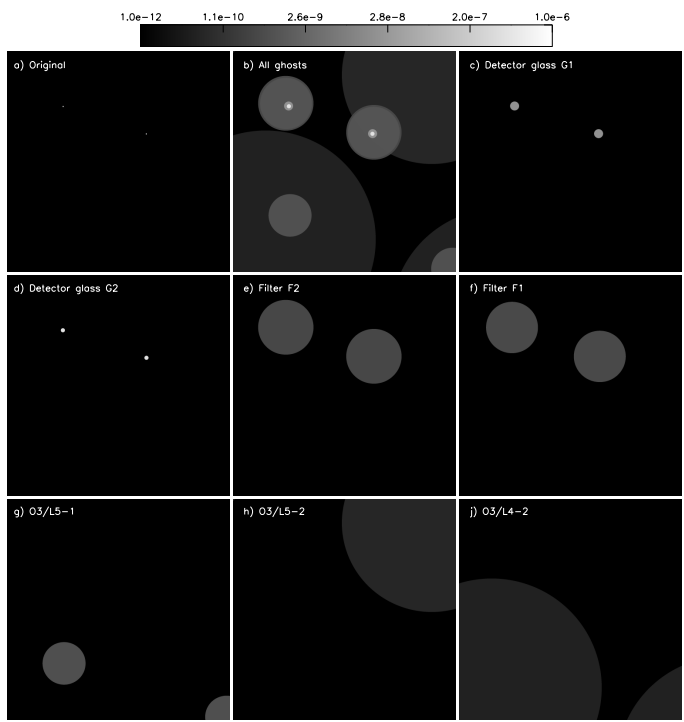}}
    \caption{Ghost images produced for an input image with two bright pixels (up-right and up-left from the center) representing two
      plane-parallel beams tilted at $0.28^\circ$ and $0.56^\circ$. The panels denote: a) the input image; b) full ghost image; c) ghost image
      produced by the front surface of the detector glass; d) by the rear surface of the detector glass; e) by the rear filter surface; f)
      by the front filter surface; g) by the rear surface of the O3/L5 lens; h) by the front surface of the O3/L5 lens; j) by the rear
      surface of the O3/L4 lens. Logarithmic color scale (shown in the top of the figure) is normalized to the intensities of the input beams.}
    \label{ghost_plane_parallel}
  \end{figure}

  We implemented this algorithm in Fortran and fed it with all the geometrical characteristics and reflectivities. In
  Fig.~\ref{ghost_plane_parallel} we show examples of ghost images calculated with the model for an input image with two bright pixels that
  represent two plane-parallel beams tilted at angles $0.28^\circ$ and $0.56^\circ$. 

  Comparison of the model presented in Fig.~\ref{ghost_plane_parallel}b with the ghost image calculated in the Zemax non-sequential regime
  (Fig.~\ref{ghost-nature}) reveals a perfect correspondence for most of the ghost contributions. We found a small mismatch of the O3/L5-1
  contributions, which were slightly shifted outwards from the center and had a small ellipticity in the case of Zemax non-sequential model. We believe the
  mismatch is due to the fact that geometrical characteristics of the ghosts were obtained in Zemax in a paraxial approximation. A future
    update of the model will take this discrepancy into account. 

  The procedure for calculating the ghost images from a particular surface very much resembles convolution. However, since most of the ghosts
  are not co-centered with the original pixel, the procedure cannot be immediately implemented as a convolution. This essentially precludes implementing
  the algorithm in an interpreted programming language (such as IDL, Python or MATLAB), since two nested loops would run
  for a very long time for a $2048 \times 2048$ image. Instead, we use a compiling programming language (Fortran) along with OpenMP
  parallelization. The full procedure takes up to 10~min for an input  $2048 \times 2048$ image on a computer with 40 Xeon e5-2580 cores.

\section{Ghost image for a realistic coronal image and the ghost removal procedure}
  \label{realistic-sec}
\subsection{Ghost image for a synthetic coronal image}
  \label{synthetic-sec}
\begin{figure*}
    \sidecaption
    \includegraphics[width=12cm]{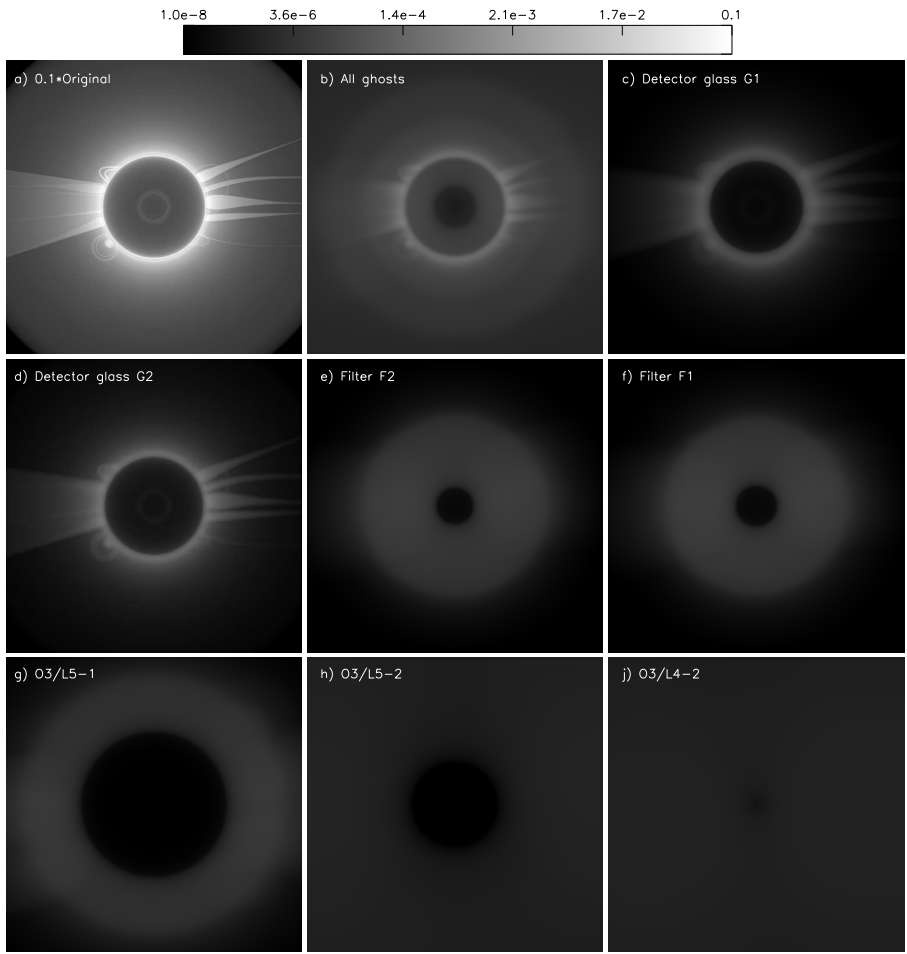}
    \caption{Ghost images for the synthetic coronal image. Panel (a) shows the input image with coronal and diffracted light superposed, other
      panels show different ghosts and have the same meaning as in Fig.~\ref{ghost_plane_parallel}.
     Color scale is normalized to the intensity of the input image. The intensity in panel (a) is divided by 10 for the
     visualisation purposes.}
    \label{ghost_images}
  \end{figure*}

  The mechanism of ghost light formation and the images presented in Fig.~\ref{ghost_plane_parallel} do not tell us how the effect will
  degrade real observational data. In order to understand this, we calculated the ghost image for a synthetic coronal image. Such an
  image is presented in Fig.~\ref{ghost_images}a. The synthetic image consists of two components: a synthetic coronal scene, and a diffracted
  light scene. Inclusion of diffracted light in the input image is important because the diffraction has significant intensity in the inner
  part of the FOV, and as a result this light could have a significant impact on the outer part. 

  We created the coronal scene in the following way: initially we created a mask with an appropriate spatial scale and FOV and populated it with
  various structures: equatorial streamers, equatorial and polar quiet Sun regions, and finally the region that corresponds to the occulter.
  After that we calculated the intensity in each pixel based on the height above the solar limb and the type of coronal structure to which it
  belongs.  The radial dependencies of intensities are taken from  \citet{1976asqu.book.....A}, and have typical values from $10^{-5}$ MSB
  at heights $\sim 1.01\mathrm{R}_\sun$ (equatorial streamers) to $10^{-10}$ MSB at heights $\sim 3\mathrm{R}_\sun$ (polar region during solar minimum). The
  coronal scene is vignetted in the inner zone due to vignetting produced by the IO. The diffracted light scene was calculated following the
  algorithm of \citet{Shestov2018} for the symmetrical ASPIICS configuration and the IO size $r_{IO}=1.662$~mm. The diffraction pattern has a
  bright ring coinciding with the size of the IO, and a smaller bright ring that corresponds to the internal opening in the IO.

  The resulting synthetic image does not represent real corona in terms of variety of observed structures. However, it has the correct
  dynamic range and reasonable radial decrease of intensity. High contrast between structures makes it convenient for an investigation of the optical properties of the telescope.
  
  Fig.~\ref{ghost_images} compares the synthetic coronal image (panel a) and calculated ghosts (same notation as in
  Fig.~\ref{ghost_plane_parallel}). It confirms the preliminary result that ghosts redistribute the light from the inner corona to outer
  FOV. In particular, the effects produced by O3/L5 and O3/L4-2 introduces a significant contribution to the outer corona. The effect of the
  detector glass looks like a smeared version of the original image, thus it may effectively broaden the point spread function of the telescope. It is
  interesting to note that the ghost images formed by the filter and lenses lose any internal structure, regardless of the fact that
  the input image had structures with very high contrast.  This fact may open the possibility of calculating ghost images with a simplified approach
  based on convolutions, which may speed-up the on-ground routine processing. Another possibility can be to use a particular coronal image
  and calculated ghost image for a set of subsequent observations, in which the corona does not change significantly.

\begin{figure*}
    \centering
    \includegraphics[width=17cm]{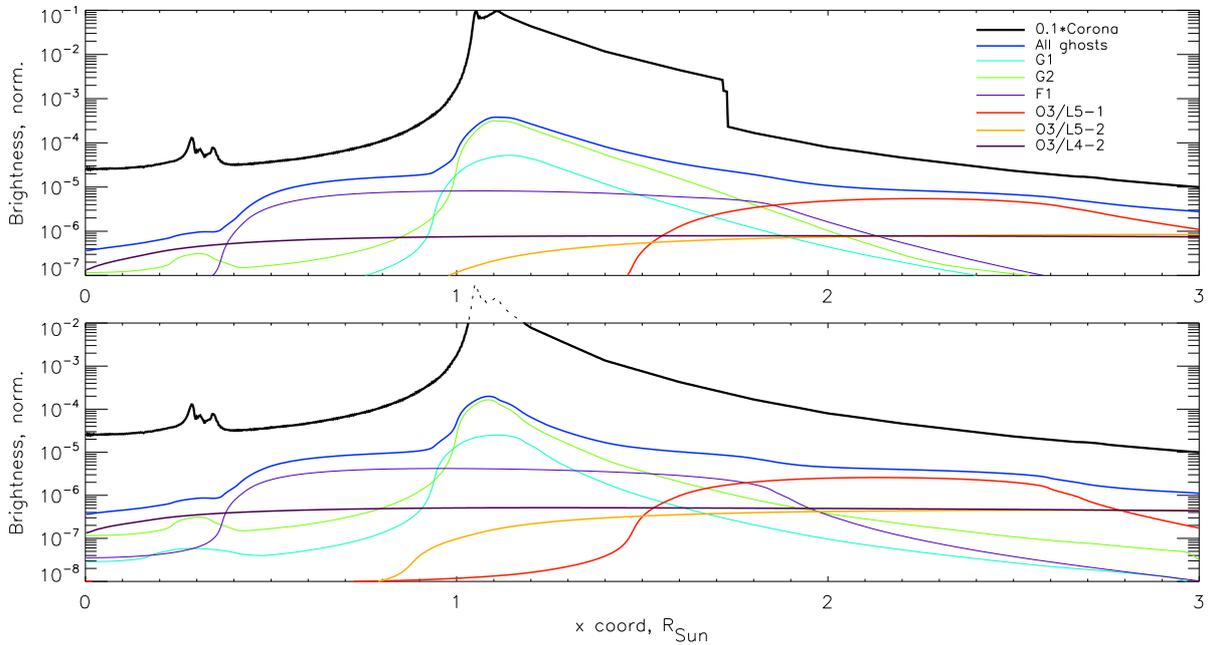}
    \caption{Radial profiles of the intensity of the synthetic coronal image and various ghost contributions. The upper panel corresponds to the
      westward equatorial direction in Fig.~\ref{ghost_images}a that crosses the equatorial streamer and the quiet Sun region, the lower panel corresponds to
      the northward equatorial direction in Fig.~\ref{ghost_images}a that crosses a polar quiet Sun region. In both panels black solid lines
      show the corona and diffraction (intensity is divided by 10), dark blue lines show the sum of all ghost contribution, other colored
      lines show contributions from various optical surfaces.}
    \label{profiles}
  \end{figure*}

  We compare radial profiles of the intensities of the input image and various ghost contributions in Fig.~\ref{profiles}. In the upper
  panel the profiles are measured along the horizontal line (westward from the disk center in Fig.~\ref{ghost_images}a) that crosses a
  streamer and a quiet equatorial region. In the lower panel the profiles are measured along the vertical line (northward from the disk
  center in Fig.~\ref{ghost_images}a) that crosses quiet polar region. 
  The profiles in the upper panel represent the worst case, because the weak outer corona is superposed with the light from bright inner coronal
  structures. The main ghost contributors for the outer corona are the O3/L5 and O3/L4 lenses. In the inner corona the main contribution
  is from G2 -- the rear surface of the detector glass. At the heights above $\sim 2 \mathrm{R}_\sun$ the ghost intensity amounts to 3\% of the
  coronal signal. This effect is achieved by the introduction of the ND50\% filter, which reduces intensity of ghosts by a factor four with
  respect to the outer corona.  
  
  
\subsection{Removal of the ghost light}
  \label{removal-sec}
\begin{figure*}
    \sidecaption
    \includegraphics[width=12cm]{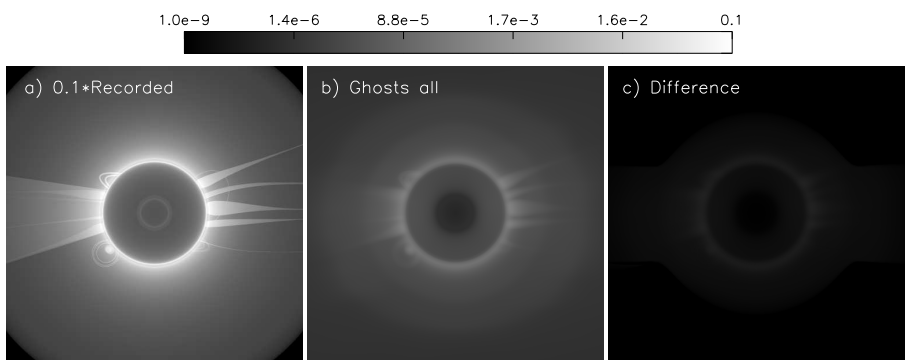}
    \caption{Analysis of the removal procedure for the synthetic coronal image. Left panel: recorded image $\mathbf{R} =
    \mathbf{I}+\mathbf{G}(\mathbf{I})$; middle panel: ghost image $\mathbf{G}(\mathbf{R})$; right panel: difference
    $\mathbf{I}-\mathbf{C}$ between the initial image and the cleared image. Color scale is normalized to the intensity of the recorded
    image. The intensity in panel (a) is divided by 10 for the visualisation purposes.}
    \label{removal}
  \end{figure*}

  The algorithm described in Sect.~\ref{model-sec} provides the ghost image for a highly idealized input image, i.e. that does not contain ghost
  light so far. In reality, the detector of the ASPIICS telescope will record images that contain both the image itself and the ghost light in a single observation. The
  problem of determining the ideal input image given the degraded image is not straightforward. However, since the intensity of the ghost image
  is significantly smaller than that of the input image, we can use the following approach. Let $\mathbf{I}$ be an ideal input image,
  $\mathbf{G}()$ is a function to calculate the ghost image, and $\mathbf{R}$ is the recorded image: $\mathbf{R} =
  \mathbf{I}+\mathbf{G}(\mathbf{I})$.  We apply the ghost function $\mathbf{G}()$ to both parts of the equation and obtain:
  \begin{equation}
    \mathbf{G}(\mathbf{R}) = \mathbf{G}\left(\mathbf{I}+\mathbf{G}(\mathbf{I})\right) = \mathbf{G}(\mathbf{I}) +
    \mathbf{G}\left(\mathbf{G}(\mathbf{I})\right)
  \end{equation}  
  as  $\mathbf{G}()$ is obviously a linear function. Since in every point $\mathbf{G}(\mathbf{I})_\mathrm{xy} < \mathbf{I}_\mathrm{xy}$,
  we obtain that $\mathbf{G}(\mathbf{G}(\mathbf{I}))_\mathrm{xy} < \mathbf{G}(\mathbf{I})_\mathrm{xy}$, and thus we can assume that:
  \begin{equation}
    \mathbf{G}(\mathbf{R}) \approx \mathbf{G}(\mathbf{I})
    \label{approx-eq}
  \end{equation}
  The validity of the approximation can be verified numerically. We calculated both the ghost image $\mathbf{G}(\mathbf{I})$ and recorded
  image $\mathbf{R}$ (Fig.~\ref{removal}a) for the synthetic input image from Fig.~\ref{ghost_images}a. Based on the degraded image we calculated
  ghosts as $\mathbf{G}(\mathbf{R})$ (Fig.~\ref{removal}b) and a cleared image $\mathbf{C}=\mathbf{R}-\mathbf{G}(\mathbf{R})$. We show
  the difference between $\mathbf{I}$ and $\mathbf{C}$ in Fig.~\ref{removal}c. The difference, i.e. the remaining ghost signal left after
  the removal procedure, is significantly weaker than the original ghost image  $\mathbf{G}(\mathbf{R})$ and on average amounts to $10^{-5}$
  of the input. Following the algorithm presented in Sect.~\ref{model-sec} we calculated ghosts of the ghosts
  $\mathbf{G}(\mathbf{G}(\mathbf{I}))$. The intensity turned out to be $\sim 2-3$ orders of magnitude less than the original
  ghost image $\mathbf{G}(\mathbf{I})$, and its spatial pattern resembled the pattern of original ghosts pattern. Thus the key assumption
  $\mathbf{G}(\mathbf{G}(\mathbf{I}))_\mathrm{xy}< \mathbf{G}(\mathbf{I})_\mathrm{xy}$ is valid with a good accuracy.
  
  This result justifies Eq.~(\ref{approx-eq}) and the validity of the proposed algorithm. It also confirms that the effect of high-order
  ghosts is negligible (see Sect.~\ref{ghosts-basic-sec}).
  
\subsection{Ghost image for a real coronal image}
    \label{eclipse-sec}
    In this section we test the algorithm described above using a real coronal image recorded during a total solar eclipse. 
   
    These eclipse observations were obtained during the 1999 August 11 eclipse in R\^amnicu V\^alcea, Romania, in an experiment originally
    designed to produce a high-quality, fully calibrated image that could be used to constrain stray light in images from the LASCO C1
    coronagraph \citep{Brueckner1995}. In spite of the loss of C1 after the SOHO failure in 1998, which rendered that goal moot during the
    eclipse, the observation proceeded and yielded a high-quality image that captures the 530.3~nm Fe XV so-called coronal green line across
    its complete dynamic range \citep{Seaton2001}.

    These observations used a purpose-built baffled simple lens telescope (to minimize the risk of internal reflection) with a 5~cm
    diameter, 300~mm focal length, and focal ratio of f/6. Spectral selection was achieved using a Dayster 530.3~nm-centered filter with a
    0.36~nm passband and a peak transmission of 50\%. The camera was a 14-bit Photometrics PM512 CCD with a 20~$\mu$m pixels, a
    $512\times512$ field of view, with a measured platescale of 14.03 arcsec/pixel. The CCD gain was 16 e$^-$/DN. To cover the entire dynamic
    range of the corona, which ranges more than three orders of magnitude between 1 and $4\mathrm{R}_\sun$, with adequate signal-to-noise and no
    saturation, we used multiple exposure times ranging from 0.25 to 32~s. However, we discarded the 32-s exposure before constructing a final composite
    because blooming from extreme saturation in the inner corona obscured a large fraction of the field of view. 

    Individual images were calibrated using flats and darks obtained just before and after the eclipse, and were exposure-normalized.
    The final image is a composite of nine separate exposures, each masked to exclude regions of saturation
    and regions with poor signal-to-noise. Because, for this analysis, we seek a smooth image with minimal noise, it is important to
    suppress temporal noise as much as possible. To help achieve this each of the individual input images is treated using a local filter
    that replaces any pixel that exceeds the range of its eight neighbors with the local maximum (or minimum, where appropriate).
    The individual masked images are then combined into a single high-dynamic-range (HDR) composite by computing the median value of the
    inputs for each pixel. 
    
    The absolute radiometric calibration of the HDR composite result is achieved by using a separate set of reference images of the
    uneclipsed Sun. From these we compute the radiance per pixel of the full Sun.

    The rate of fall-off of the green-line corona in the resulting image is roughly consistent with observations of the green line at the
    1981 total eclipse by \citet{1997ASIC..494..159K}, which show a decrease of roughly three orders of magnitude between heights of 1 and
    $2\mathrm{R}_\sun$. It is worth noting that small amount of residual signal is present in the darkest areas of the image, likely the result of
    scattered light inside the telescope, but it is significantly less than 1 part in 1000 relative to the radiances recorded in the inner
    corona, and thus is only significant at large heights. Likewise, areas of blooming in the longest exposures rendered a some faint regions
    nonetheless unusable for compositing, which led to a small increase in noise in some areas of the image. Again, the noise is
    insignificant relative to the overall observed brightness in much of the corona.

    This final calibrated image is then resampled to the correct ASPIICS plate scale and resolution and the data rescaled according to the effective area
    of the ASPIICS instrument. 

\begin{figure*}
      \sidecaption
      \includegraphics[width=12cm]{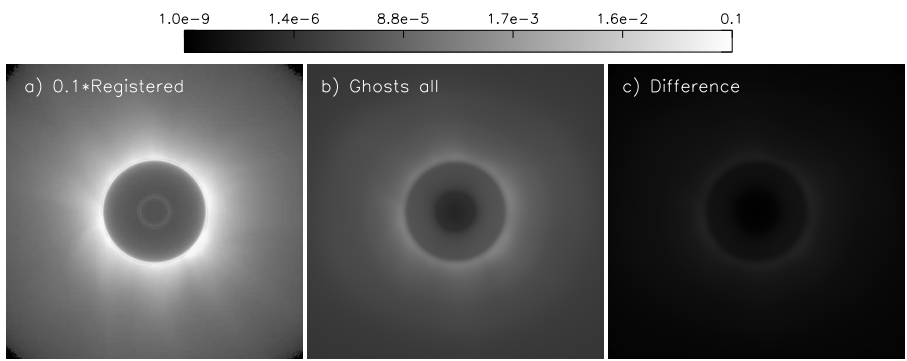}
      \caption{Analysis of the removal procedure for the coronal image obtained during a total solar eclipse. Panels have the same meaning as in
	Fig.~\ref{removal}.      }
      \label{eclipse-removal}
    \end{figure*}

    As before, we add the diffracted light calculated for the IO with $r_{IO}=1.662$~mm and corresponding vignetting of the corona. In
    Fig.~\ref{eclipse-removal} we compare the registered image, the ghost image and the difference with the original.

    As in the case with the synthetic image, the removal algorithm performs well and the residual, uncorrected signal amounts to $10^{-5}$ of the
    original corona.

\section{Comparison with other sources of stray light}
  \label{comparison-sec}
  In this section we compare the brightnesses of the corona with that of the various stray light contributors: diffracted light, ghost light and
  scattered light. We take intensity of the corona and the diffracted light on the detector as in Sect.~\ref{synthetic-sec}
  ($r_{IO}=1.662$~mm, symmetrical ASPIICS configuration, corona is vignetted by the IO). We calculate the ghost and
  scattering images individually for the coronal and for the diffracted light scenes.
  
  The calculation of the scattered light image is based on the approach and parameters presented in \citet{Galy2018}, however here we
  calculate scattering for a realistic coronal scene. To achieve this, we consider the effect on the detector from scattering at a
  particular lens as a convolution with some kernel. We obtain the kernel from the raytracing modeling and then convolve the input image with the
  kernel. Since the scattering on lens surfaces is rather small, with typical values of total integrated scatter (TIS;
  \citealt{doi:10.1117/1.OE.51.1.013402}) being $10^{-4}$ of the incoming radiation, we consider scattering on each lens independently.
  Such an approach corresponds to consideration of the first order scatter, in which every individual ray is allowed to scatter no more than once. 

\begin{figure}
    \includegraphics[width=8cm]{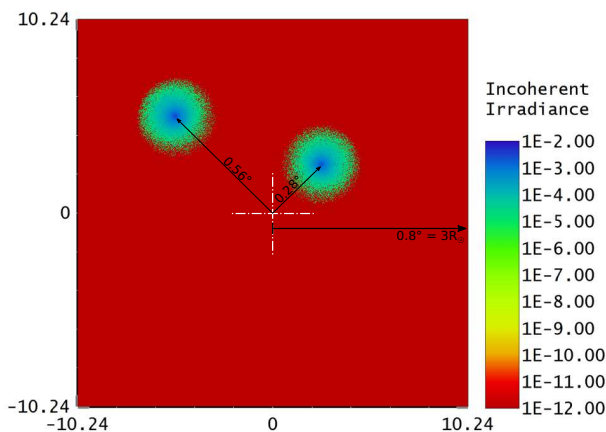}
    \caption{Image on the detector produced in Zemax OpticStudio in non-sequential regime with scattering on the O2 lens for the two input
      beams tilted at $0.28^\circ$ and $0.56^\circ$. The logarithmic color scale corresponds to irradiance and is normalized to the peak
      irradiance of the scattered images.}
    \label{scattering}
  \end{figure}

  Lens scattering characteristics were inferred from the atomic force microscopy (AFM) measurements. Surface root-mean-square
  (RMS) micro-roughness of the lens surfaces ranged from 0.6~nm (both lenses of the primary objective) to 3.0~nm (O3/L2 -- second lens of
  the relay lens). Based on these parameters and using an ABg representation \citep{Pfisterer2011} of the bi-direction scattering distribution
  function (BSDF) in the non-sequential regime of Zemax OpticStudio, we calculated a response on the detector produced for two incoming
  plane-parallel beams. The strongest scattered signal was produced by the O2 lens (see Fig.~\ref{scattering}) and O3/L1 lens. For these
  lenses the following parameters were applied: RMS micro-roughness $\sigma \sim 2.6$~nm, $\mathrm{TIS}=3 \cdot 10^{-4}$, and fitted ABg
  parameters $A=3\cdot 10^{-5}$, $B=10^{-3}$, $g=1.5$. Scattering produced by the rest of the optics had even smaller
  intensity and produced diffuse uniform images. 

  The image presented in Fig.~\ref{scattering} shows scattering produced by the O2 lens for the two input plane-parallel beams
  titled at angles $0.28^\circ$ and $0.56^\circ$. We fit each kernel with the analytical formula $I(r)=I_0\cdot10^{-r/r_0}$. The parameter
  $r_0$ was determined to be $r_0=0.5$~mm, and the amplitude $I_0$ was adjusted to produce the total intensity of $1.3\cdot 10^{-5}$ of the
  initial beam (corresponding to the Zemax result). 
  
  The O3/L1 lens produced a wide scattering pattern that we approximated as  $I(r)=I_0\cdot
  \exp\left(-\frac{r^2}{2w^2}\right)$. The parameters were fitted as $w = 7.0$~mm, and $I_0$ was adjusted to produce the total
  intensity of $1.05\cdot 10^{-5}$ of the initial beam.

  We used the two kernels obtained above to produce scattering for the realistic input image containing coronal and diffraction
  scenes (from Fig.~\ref{ghost_images}a).   

  Since the amount of diffracted light coming to the primary objective is significantly larger than that on the detector, we
  analyzed scattering on the PO individually. We determined scattering properties for the PO, for which we used the following
  parameters: RMS micro-roughness $\sim 1$~nm, $\mathrm{TIS}=5.5\cdot10^{-5}$, ABg parameters:  $A=5.5\cdot10^{-6}$, $B=10^{-3}$, $g=1.5$.
  Raytracing showed that the scattering was mainly determined by the O1/L1 lens, and the pattern on the detector was fitted as
  $I(r)=I_0\cdot \exp\left(-\frac{r^2}{2w^2}\right)$, where $w=11.0$~mm, and the amplitude $I_0$ was adjusted to produce the total
  inensity $9.08\cdot 10^{-7}$ of the initial beam. This kernel was convolved with the radial profile of the diffracted light corresponding
  to the one presented in Fig.~6 in \citet{2017A&A...599A...2R}.
  
\begin{figure}
    \resizebox{\hsize}{!}{\includegraphics{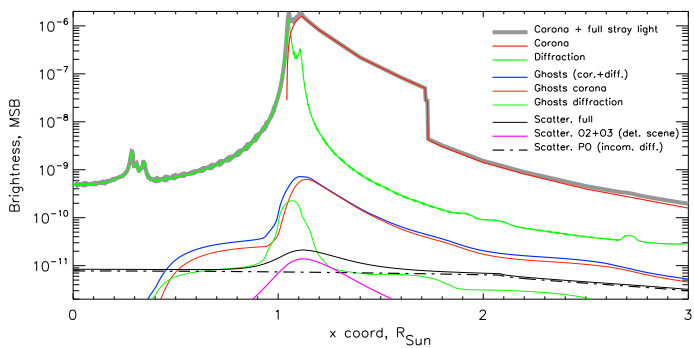}}
    \caption{Radial profiles of the intensity of the coronal light, the diffraction, ghost, and scattered light contributions. The thick red 
      line corresponds to the corona, the thick green line corresponds to the diffraction, and the thick gray line corresponds to the full signal
      on the detector. The dark blue line represents ghosts of the full image, the thin red and green lines are contributions due to coronal
      ghosts and diffracted light ghosts. The thin black line shows full scattering light, the dot-dashed line and magenta line show
      contribution from the primary objective (dot-dashed) and the O2 and O3 lenses (magenta line). }
    \label{corona_vs_diff}
  \end{figure}

  The results are presented in Fig.~\ref{corona_vs_diff} showing all the contributors of stray light. The coronal profile (thick red line)
  corresponds to the direction westward from the disk center in Fig.~\ref{ghost_images}a and crosses a streamer and a quiet region. The
  ghost light was calculated for the corona and the diffracted light (which includes only the intensity on the detector, since the intensity of
  ghosts produced by O1 is negligible). The scattered light contains components produced by O1 and strong incoming diffracted light, and O2
  and O3 component produced by both coronal light and the diffracted light signal. 
  
  Fig.~\ref{corona_vs_diff} demonstrates that the scattering has a smaller contribution than the ghost reflection and diffraction. Further,
  the contribution of ghosts is smaller than that of the diffracted light; we note, however, that the diffracted light given here is
  calculated for an ``average'' case -- for the symmetrical configuration (tilt of the telescope would increase the intensity of the
  diffraction ring in one side, see \citet[][]{Shestov2018}) and a rather small $r_{IO}=1.662$~mm. Comparison of the green and red curves
  demonstrates that the ghost contribution due to diffracted light will be smaller than the corona. The major contribution of the scattered
  light is produced by O1.  This contribution is rather uniform across the detector, whereas the smaller contribution from O2 and O3 lenses
  has a relatively more intense central part. 

\section{Discussion and Conclusions}
  \label{discussion-sec}
  Here, we analysed sources of the ghost light on the ASPIICS detector. We showed that due to the relatively high reflectivity of the
  detector (15\%) and low reflectivity of the AR coatings ($\sim0.3$\%) the main ghosts are produced by the backreflection from the detector
  and the neighbouring optical surfaces. These contributions are relatively easy to model because of the simple geometrical behaviour and
  absence of vignetting of the ghost beams inside the telescope. Redistribution of the light from the inner corona to the
  outer corona caused by the ghost reflections makes the effect important for observation of the outer corona. Currently, with the presence
  of the ND50\% filter, the ghost signal amounts to 3\% of the coronal signal.

  We created a model that calculates the ghost image for a given input image.  The major ghost contributors that are not taken into account
  in the current model are created by the filter and the detector glass and the other optical surfaces. The ghost image
  calculated for a synthetic coronal image resembles a blurred coronal image. This fact opens the possibility of adjusting the model
  empirically based on the laboratory measurements and calibrations. Additionally, it may allow the development of a simplified algorithm for the ghost
  calculation. 

  We showed that the ghost image can be calculated based on the image recorded at the detector, i.e. an image that contains the ghost signal. The removal
  procedure works very efficiently, and after the removal, the relative contribution of residual ghost light amounts to $10^{-5}$ of the local
  coronal signal. The procedure yields good results even with a stronger contribution of ghosts, e.g. with the removed detector glass
  and the ND50\% filter (see Appendix~\ref{no-detector-glass-sec}). This fact raises the question of the necessity of the detector glass and the ND50\%
  filter, as, besides reducing the ghost contributions, the glass introduces its own ghosts. Any possible tilt of the glass
  (even very small) may further complicate the situation.

  For a comparison with other stray light sources, we took the diffraction calculated from \citet{2017A&A...599A...2R,Shestov2018}, and ABg scattering on
  the lens surfaces. To model the scattering, we took the worst-case ABg parameters with $\mathrm{TIS}=3\cdot10^{-4}$ and studied individually
  the effect on the detector produced by every lens. The main contributors are the O2 and O3/L1 lenses. We calculated the scattering of coronal and
  diffracted light due to these lenses. We also analysed scattering of the intense incoming diffracted light (whose total intensity is 3 orders of
  magnitude larger than the intensity coming to the detector) by the primary objective without occultation by the IO. We believe this approach
  is sufficient to obtain a qualitative result. 

  The analysis shows that the major contributor to the stray light is diffraction. Its intensity is at least one order of magnitude larger
  than ghost light or scattering, and amounts to 10\% of the observed coronal signal in the outer corona region. The intensity of the ghost
  light is almost 10 times smaller in the outer region, and the intensity of the scattered light (calculated in the raytracing/BSDF
  approach) is smaller than the intensity of the ghost light.  The small contribution of the ghost light in comparison to diffraction
  further argues for the non-necessity of the detector glass. The use of the detector glass inevitably introduces an additional ghost
  component, and using of the ND50\% reduces the total throughput of the telescope. This may be especially important for observations of
  polarized light. Recent consideration of diffraction and scattering together \citep{Rougeot2018b} shows that, due to scattering, the
  diffracted light increases its intensity in the outer corona (this is not taken into account in other models). This possibility even
  further increases the importance of the diffracted light and reduces relative contributions of other sources.

  We showed that the ghost light produced by the primary objective is relatively weak due to ASPIICS's optical design. Thus we conclude that the
  Lyot spot (used in LASCO C2 and LASCO C3) is not necessary in ASPIICS.
 
  The unique feature of the ASPIICS coronagraph provided by the FF capability is the observation of extremely low corona, starting from
  heights as low as $\sim 1.08\mathrm{R}_\sun$. Such observations are essentially impossible to obtain with other types of
  externally occulted space coronagraphs, because of significant stray light and vignetting. In fact, the requirements to
  reduce stray light and simultaneously minimize the inner observational height effectively counteract each other, as the diffraction in the
  internal zone decreases only with the increase of the occultation and vignetting of the corona \citep{2017A&A...599A...2R}. Additional
  stray light on the detector not only increases possible noise in each individual pixel due to photon noise, but, more importantly, alters
  the inferred photometry. Photon noise (as well other types of noise) can be reduced by using two sequential exposures or averaging over
  several pixels, whereas systematic overestimation of measured intensity due to stray light may result in misinterpretation of
  observational data. Careful laboratory measurements of the ghost light and scattered light performance are necessary for the development
  of successful on-ground data processing. An especially important role is played by validation of the correctness of the adopted models,
  i.e. confirming the absence of other sources of stray light, correct identification of the major ghost sources, the applied scattering
  model, etc. 

  Current models for the stray light, which take into account diffraction \citep{2017A&A...599A...2R,Shestov2018}, ghost reflections
  \citep[the present analysis and ][]{Galy2018} and scattering \citep{Rougeot2018b}, show that the total contribution of the stray light will
  be less than 10\% of the coronal signal almost in the whole FOV. This makes the ASPIICS an instrument that will provide unique
  observations of the low solar corona. 


\begin{acknowledgements}
  S.~V.~S. and A.~N.~Z thank the European Space Agency (ESA) and the Belgian Federal Science Policy Office (BELSPO) for their support in the framework of the PRODEX Programme.  
\end{acknowledgements}

\bibliographystyle{aa}
\bibliography{shestov_zhukov}

\appendix
\section{Raytracing in Zemax OpticStudio}
\label{zemax}
  In order to analyse properties of ghost images, we use a raytracing software package -- Zemax
  OpticStudio\footnote{\url{https://www.zemax.com/products/opticstudio}} in various regimes.
  In a typical user scenario the user sets up the optical layout by specifying the relative positions, radii of curvatures, thicknesses and
  diameters of the lenses, as well as glass material and possibly other properties (e.g. coating). The user also provides the input beam (one or
  several) by specifying its field angle. Then a set of rays are launched by Zemax. The rays have the same field angle but are shifted with
  respect to each other perpendicularly to the optical axis to fill the whole entrance aperture of the optical system. Zemax traces every ray,
  calculating (within numerical accuracy) the coordinates of the intersection of the ray with a particular lens' surface, taking into
  account all geometrical factors. Then the angles of incidence and transmission are calculated, and the ray is traced further until the
  next surface and so on, until it ultimately reaches the detector. The final position of all the incoming rays produces the point spread
  function (PSF) of the telescope. This regime of Zemax operation is called sequential, because every individual ray passes all the lenses
  in the order of their $z$-coordinate and it never gets splitted. 
  
  In sequential analysis, Zemax also provides the user with the first-order parameters of the optical system, such as effective focal
  length $f$, working $f/\#$ number, magnification, etc. These parameters are obtained in the paraxial regime (i.e. limit of very small
  ray angles and heights), however these parameters are not applicable when the full geometry model (not paraxial) is used, i.e. not
  applicable already in the sequential regime. 
  
  In its non-sequential regime, Zemax allows every particular ray to be splitted on a lens surface, giving rise to additional ghost or
  scattered rays. The newly introduced rays (their actual number depends on many factors) carry a small amount of intensity of the mother ray
  in accordance with reflective or scattering properties of the surface. Afterwards all the rays are considered in a similar
  manner and are raytraced further, however the information about every ray origin is preserved and can be used afterwards. In this regime no
  rule of sequential propagation through lenses is applied.  Since during consideration of ghost light and scattering the number of new rays increases as
  an avalanche with the increase of number of optical surfaces, and their intensity decreases dramatically with the number of
  backrefletions/scattering events, various limitations (like minimal intensity or maximal number of splits) are used to keep the number of
  rays under analysis reasonable. In particular, we had to fine-tune some parameters of Zemax to be able to reveal all the ghost rays (from
  all the surfaces) on the detector. During the investigation of scattering we were able to consider scattering on lenses individually, i.e.
  initially first lens, then second lens, etc. Turning on scattering on all the surfaces gave nonphysical results, as the scattered
  rays from the former surfaces were omitted. 
 
  There exists a possibility of analysing ghost images in Zemax in paraxial regime. We did not find an exact explanation of the feature
  in the Zemax documentation, but we believe it is done as follows: to investigate the ghosts formed by two particular surfaces, Zemax
  automatically duplicates the given part of the optical system and considers paraxial ray propagation within the modified optical layout.
  Within the paraxial approximation such parameters as effective focal length, position of exit pupil, effective $f/\#$ ratio etc. are
  calculated, however they are valid only for the modified system, i.e. for a particular ghost image. 
  
\section{Ghost light with no detector glass}
    \label{no-detector-glass-sec}
    \begin{figure*}
      \centering
      {\includegraphics[width=14cm]{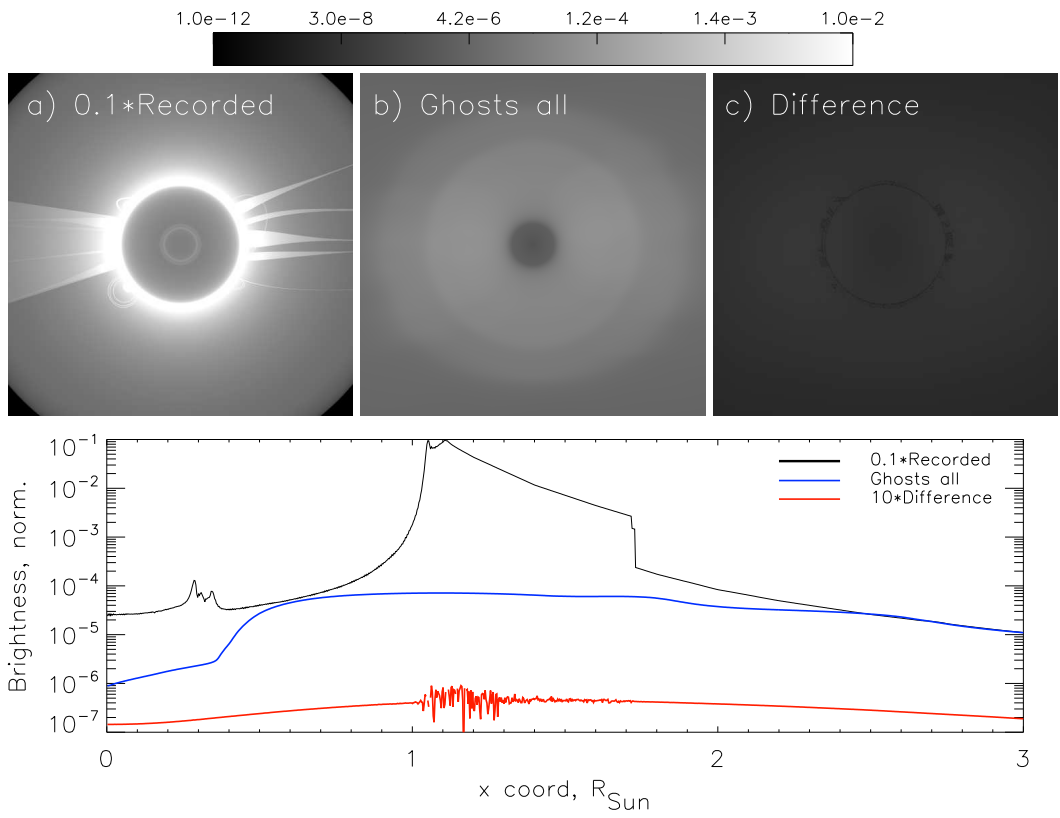}}
      \caption{Analysis of the removal procedure for the synthetic coronal image with the removed detector glass. The top panels have the same notation as in
	Fig.~\ref{removal}, the bottom panel shows intensities of $\mathbf{R}$, $\mathbf{G(R)}$ and  $\mathbf{I}-\mathbf{C}$ in the westward
	from the disk center direction.    }
      \label{no-glass-removal}
    \end{figure*}
    In this section we analyze the behaviour of the ghost light in the case we completely remove the detector glass from the optical system. Here, we
    take the synthetic coronal scene as in Sect.~\ref{removal-sec} and modify the Fortran code in order to completely remove the contribution of
    the detector glass (corresponding ghosts) and ND50\% filter (decrease of intensity of the remaining ghosts). In
    Fig.~\ref{no-glass-removal} we present a comparison of the observed image $\mathbf{R}$ (panel a), ghost image $\mathbf{G(R)}$ (panel
    b) and the difference $\mathbf{I}-\mathbf{C}$ after the removal procedure (panel c). The color scale is normalized to the recorded
    image, however the intensity in panel (a) is divided by 10 for visualization purposes. The dynamic range of the color scale is modified with respect to
    Figs.~\ref{removal},~\ref{eclipse-removal} in order to reveal weaker intensities. In the bottom panel we plot horizontal profiles of the
    registered image (black line; factor 0.1), ghost image (blue line) and the difference (red line; factor 10).

    Removing the detector glass causes the relative intensity of ghosts to increases by a factor $\approx 4$, as the ND50\% filter is no longer present to reduce ghost 
    contributions any more. In this case, the intensity of the ghosts amounts to 10\% of the corona at heights $>2.2\mathrm{R}_\sun$. However, after
    the removal procedure the remaining signal has a relative amplitude $<10^{-4}$ and has a more uniform structure. We attribute this
    to the modified structure of the ghost image, as the bright and relatively contrasted contributions
    from the detector glass are absent.

%

\end{document}